\documentclass[12pt]{article}
\usepackage{amsfonts}

\usepackage{graphicx}
\usepackage{epsf}

\usepackage{amsmath}

\begin{document}

\catcode`@=11
\long\def\@caption#1[#2]#3{\par\addcontentsline{\csname
  ext@#1\endcsname}{#1}{\protect\numberline{\csname
  the#1\endcsname}{\ignorespaces #2}}\begingroup
    \small
    \@parboxrestore
    \@makecaption{\csname fnum@#1\endcsname}{\ignorespaces #3}\par
  \endgroup}
\catcode`@=12
\newcommand{\newc}{\newcommand}
\newc{\rem}[1]{{\bf #1}}
\def\ltap{\ \raise.3ex\hbox{$<$\kern-.75em\lower1ex\hbox{$\sim$}}\ }
\def\gtap{\ \raise.3ex\hbox{$>$\kern-.75em\lower1ex\hbox{$\sim$}}\ }
\def\gl{\ \raise.5ex\hbox{$>$}\kern-.8em\lower.5ex\hbox{$<$}\ }
\def\roughly#1{\raise.3ex\hbox{$#1$\kern-.75em\lower1ex\hbox{$\sim$}}}
\newc{\gsim}{\lower.7ex\hbox{$\;\stackrel{\textstyle>}{\sim}\;$}}
\newc{\lsim}{\lower.7ex\hbox{$\;\stackrel{\textstyle<}{\sim}\;$}}
\newc{\gev}{\,{\rm GeV}}
\newc{\mev}{\,{\rm MeV}}
\newc{\ev}{\,{\rm eV}}
\newc{\kev}{\,{\rm keV}}
\newc{\tev}{\,{\rm TeV}}
\def\det{\mathop{\rm det}}
\def\tr{\mathop{\rm tr}}
\def\Tr{\mathop{\rm Tr}}
\def\Im{\mathop{\rm Im}}
\def\Re{\mathop{\rm Re}}
\def\bR{\mathop{\bf R}}
\def\bC{\mathop{\bf C}}

\newcommand{\Z}{Z}

\def\lie{\mathop{\hbox{\it\$}}} 
\newc{\sw}{s_W}
\newc{\cw}{c_W}
\newc{\swsq}{s^2_W}
\newc{\cwsq}{c^2_W}
\newc{\mgrav}{m_{3/2}}
\newc{\mz}{M_Z}
\newc{\mpl}{M_{pl}}
\def\ux{U(1)$_X$}
\def\beq{\begin{equation}}
\def\eeq{\end{equation}}
\def\bea{\begin{eqnarray}}
\def\eea{\end{eqnarray}}
\def\bi{\begin{itemize}}
\def\ei{\end{itemize}}
\def\benum{\begin{enumerate}}
\def\eenum{\end{enumerate}}

%
\def\boxeqn#1{\vcenter{\vbox{\hrule\hbox{\vrule\kern3pt\vbox{\kern3pt
\hbox{${\displaystyle #1}$}\kern3pt}\kern3pt\vrule}\hrule}}}
%

\def\qed#1#2{\vcenter{\hrule \hbox{\vrule height#2in
\kern#1in \vrule} \hrule}}
\def\half{{\textstyle{1\over2}}} 
\newc{\ie}{{\it i.e.}}          \newc{\etal}{{\it et al.}}
\newc{\eg}{{\it e.g.}}          \newc{\etc}{{\it etc.}}
\newc{\cf}{{\it c.f.}}
\def\CAG{{\cal A/\cal G}}
\def\CA{{\cal A}} \def\CB{{\cal B}} \def\CC{{\cal C}} \def\CD{{\cal D}}
\def\CE{{\cal E}} \def\CF{{\cal F}} \def\CG{{\cal G}} \def\CH{{\cal H}}
\def\CI{{\cal I}} \def\CJ{{\cal J}} \def\CK{{\cal K}} \def\CL{{\cal L}}
\def\CM{{\cal M}} \def\CN{{\cal N}} \def\CO{{\cal O}} \def\CP{{\cal P}}
\def\CQ{{\cal Q}} \def\CR{{\cal R}} \def\CS{{\cal S}} \def\CT{{\cal T}}
\def\CU{{\cal U}} \def\CV{{\cal V}} \def\CW{{\cal W}} \def\CX{{\cal X}}
\def\CY{{\cal Y}} \def\CZ{{\cal Z}}
\def\grad#1{\,\nabla\!_{{#1}}\,}
\def\gradgrad#1#2{\,\nabla\!_{{#1}}\nabla\!_{{#2}}\,}
\def\partder#1#2{{\partial #1\over\partial #2}}
\def\secder#1#2#3{{\partial^2 #1\over\partial #2 \partial #3}}
\def\slash#1{\rlap{$#1$}/} 
\def\dsl{\,\raise.15ex\hbox{/}\mkern-13.5mu D} 
\def\delsl{\raise.15ex\hbox{/}\kern-.57em\partial}
\def\Ksl{\hbox{/\kern-.6000em\rm K}}
\def\Asl{\hbox{/\kern-.6500em \rm A}}
\def\Dsl{\hbox{/\kern-.6000em\rm D}} 
\def\Qsl{\hbox{/\kern-.6000em\rm Q}}
\def\gradsl{\hbox{/\kern-.6500em$\nabla$}}

%
\let\al=\alpha
\let\be=\beta
\let\ga=\gamma
\let\Ga=\Gamma
\let\de=\delta
\let\De=\Delta
\let\ep=\varepsilon
\let\ze=\zeta
\let\ka=\kappa
\let\la=\lambda
\let\La=\Lambda
\let\del=\nabla
\let\si=\sigma
\let\Si=\Sigma
\let\th=\theta
\let\Up=\Upsilon
\let\om=\omega
\let\Om=\Omega
\def\ph{\varphi}
\newdimen\pmboffset
\pmboffset 0.022em
\def\oldpmb#1{\setbox0=\hbox{#1}%
 \copy0\kern-\wd0
 \kern\pmboffset\raise 1.732\pmboffset\copy0\kern-\wd0
 \kern\pmboffset\box0}
\def\pmb#1{\mathchoice{\oldpmb{$\displaystyle#1$}}{\oldpmb{$\textstyle#1$}}
        {\oldpmb{$\scriptstyle#1$}}{\oldpmb{$\scriptscriptstyle#1$}}}
\def\bar#1{\overline{#1}}
\def\vev#1{\left\langle #1 \right\rangle}
\def\bra#1{\left\langle #1\right|}
\def\ket#1{\left| #1\right\rangle}
\def\abs#1{\left| #1\right|}
\def\vector#1{{\vec{#1}}}
\def\inv{^{\raise.15ex\hbox{${\scriptscriptstyle -}$}\kern-.05em 1}}
\def\pr#1{#1^\prime}  
\def\lbar{{\lower.35ex\hbox{$\mathchar'26$}\mkern-10mu\lambda}} 
\def\e#1{{\rm e}^{^{\textstyle#1}}}
\def\ee#1{\times 10^{#1} }
\def\imp{~\Rightarrow}
\def\coker{\mathop{\rm coker}}
\let\p=\partial
\let\<=\langle
\let\>=\rangle
\let\ad=\dagger
\let\txt=\textstyle
\let\h=\hbox
\let\+=\uparrow
\let\-=\downarrow
\def\dot{\!\cdot\!}
\def\vfilll{\vskip 0pt plus 1filll}

\long\def\symbolfootnote[#1]#2{\begingroup%
\def\thefootnote{\fnsymbol{footnote}}\footnote[#1]{#2}\endgroup}
%
\begin{titlepage}
\begin{flushright}
{OUTP-0510P}
\end{flushright}
\vskip 2cm
\begin{center}
{\Large Signals of Inflation in a Friendly String Landscape}

\vskip 0.3cm

{\large John March-Russell\symbolfootnote[2]{E-mail: j.march-russell1@physics.ox.ac.uk} and  Francesco Riva\symbolfootnote[1]{E-mail: riva@thphys.ox.ac.uk}} \vskip 0.5in 
Rudolf Peierls Centre for Theoretical Physics\\
University of Oxford, 1 Keble Road, Oxford OX1 3NP, UK

\end{center}

\vspace*{.2in}

\begin{abstract}
Following Freivogel {\it et al} we consider inflation in a predictive (or `friendly')
region of the landscape of string vacua, as modeled by Arkani-Hamed, Dimopoulos and Kachru. 
In such a region the dimensionful coefficients of super-renormalizable operators unprotected
by symmetries, such as the vacuum energy and scalar mass-squareds are freely scanned over,
and the objects of study are anthropically or `environmentally' conditioned probability distributions
for observables.  In this context we study the statistical predictions of (inverted) hybrid inflation
models, where the properties of the inflaton are probabilistically distributed.  We derive the
resulting distributions of observables, including the deviation from flatness
$|1-\Omega|$, the spectral index of scalar cosmological perturbations $n_s$
(and its scale dependence $dn_s/d\log k$), and the ratio of tensor to scalar perturbations $r$.
The environmental bound on the curvature implies a solution to the $\eta$-problem of
inflation with the predicted distribution of $(1-n_s)$ indicating values close to
current observations.  We find a relatively low probability ($<3\%$) of `just-so' inflation
with measurable deviations from flatness. Intermediate scales of inflation are preferred in these models.
\end{abstract}

\end{titlepage}

\section{Introduction and Motivation}

Recent developments in string theory suggest that the universe we observe might be nothing more
than one member of an exponentionally large number of possible metastable de Sitter vacua -- the {\it landscape}
\cite{Bousso:2000xa,Feng:2000if,Kachru:2003aw}.  The number
of vacua is likely so large as to be able to allow the apparent fine-tuning of the
cosmological constant via weak anthropic or `environmental' arguments along the lines of
Weinberg's now famous study \cite{Weinberg:1987dv,Weinberg:1988cp}.
This would have dramatic consequences for our understanding of fundamental physics.  Indeed, if
the landscape of string theory exists, the
approach taken by physicists when studying properties of our universe will change completely,
from the search for a single vacuum state leading to the standard model of particle physics and cosmology,
to a statistical approach, in which one is interested in the likelihood of finding the
standard model among all the possible vacua of string theory conditioned on some basic `environmental'
requirements, such as the existence of atoms and collapsed galactic structures.  Such statistical
arguments can be used both to test the theory from which the landscape originates and to make
predictions in terms of conditioned relative probabilities
\cite{Weinberg:1988cp,Tegmark:1997in,Kachru:2003aw,Douglas2,Tegmark:2005dy}. 

Freivogel {\it et al} (FKRMS) \cite{FKMS} argued that the landscape also has implications for the structure
and evolution of the universe.  At the largest
scales it is eternally inflating and continually producing island universes by tunneling events. 
Indeed our own region of the universe is predicted to have experienced a number of such events,
the last tunneling event leading to a conventionally inflating universe
that eventually evolved into the region of the landscape with the almost vanishing cosmological
constant we now observe, and with some of the observed laws of physics (and standard model parameter values)
contingent on our environment.  If one accepts such an environmental solution to the cosmological 
constant problem, the most pressing question becomes: Which features of our standard
model of particle physics and cosmology are environmentally determined and which are determined by dynamics?

Recently Arkani-Hamed, Dimopoulos, and Kachru (AHDK) \cite{ADK}, suggested that we might live in a particular {\it predictive}
region of the landscape where a separation between environmentally and dynamically determined quantities
can be cleanly made.  In this region, which they named the friendly neighborhood of the landscape or
`friendly landscape' for short, the dimensionful coefficients of super-renormalizable operators unprotected by symmetries 
(such as the vacuum energy, and scalar mass-squareds in the absence of exact supersymmetry) are freely
scanned over, while dimensionless quantities or protected dimensionful quantities (such as gauge
couplings and fermion masses respectively) are essentially fixed.
This allows for both the successes of the usual `unique vacuum' approach in building the standard model, such
as gauge-coupling unification which would still be determined by dynamics, together with the success of Weinberg's
environmental approach for explaining the cosmological constant.   
In what follows we assume that our universe is located in such a predictive neighborhood of the landscape.

In this paper we explore in greater detail the implications of the friendly landscape for inflation \cite{Guth:1980zm,Albrecht:1982wi,Linde:1981mu,Linde:1983gd}, and the
observable signals that arise.
The starting point for our analysis is the simple observation
that the overall scale of inflation and the $(\rm{mass})^2$ of the inflaton both correspond to unprotected
super-renormalizable quantities that according to the sharp rules of the friendly landscape \cite{ADK} should be freely scanned
over.   Moreover, the Coleman-deLuccia tunneling
events \cite{Coleman:1980aw} that the landscape predicts are in our past
produce negative curvature Friedmann-Robertson-Walker (FRW) island universes, and
without a significant period of inflation our universe would be dominated by
curvature and will never match onto standard hot big-bang cosmology. Indeed, as discussed in \cite{Vilenkin:1996ar,Garriga:1998px,FKMS}, the environmental requirement of structure formation
puts an upper bound on the curvature which is close to the experimental bound. 
The fundamental reason for this upper bound is that both positive vacuum
energy and negative curvature inhibit structure formation.  This bound on curvature
then translates to a lower bound on the duration of inflation.  

According to the arguments of FKMRS a conservative {\it environmental} (weak anthropic)
bound on curvature obtained by requiring structure on the scales of dwarf galaxy translates
into a lower bound on the number of e-folds of inflation
\begin{equation}
N_{\rm structure} > 59.5  ,
\label{Nebound}
\end{equation}
(here it is assumed that inflation takes place at the scale $V_{end}^{1/4} \approx 10^{16}${\rm GeV}).  Requiring structure
on larger distances such as typical galaxy scales slightly strengthens this bound due to the logarithmic
change of $\delta\rho/\rho$ for sub-horizon scales before radiation-matter equality. 
The bound Eq.(\ref{Nebound}) is remarkably close to the present {\it observational} bound on curvature for a $k=-1$
universe which translated into a duration of inflation reads\footnote{In deriving the observational
bound Eq.(\ref{Nobsbound}) FKMRS used 
$\Omega_{\rm total} > 0.98$, 2$\sigma$ away from the WMAP result $1.02 \pm 0.02$
\cite{Bennett:2003bz}, and 1$\sigma$ off the result of large-scale structure surveys $0.99 \pm 0.01$ 
\cite{Eisenstein:2005su}.}
\begin{equation}
N_{\rm observation} > 62.0  .
\label{Nobsbound}
\end{equation}

We show that applying the rules of the friendly landscape to the parameters of realistic inflationary models, together with the anthropic curvature bound, Eq.(\ref{Nebound}), leads to some striking statistical predictions for measurable quantities. 
The observables that we consider are the deviation from flatness $|1-\Omega|$, the spectral index of cosmological
perturbations $n_s$ (and its scale dependence $dn_s/d\log k$), and
the ratio of tensor to scalar perturbations $r$, all of which can be expressed in terms of just a
small number of parameters: the slow-roll parameters $\epsilon$ and $\eta$, the number of e-foldings $N$
and the amplitude $\delta_H=\delta\rho/\rho$  (following Weinberg \cite{Weinberg:1987dv}
we will use as an input the measured value of $\delta_H$)\footnote{The possibility of allowing $\delta_H$ to vary, has been discussed in \cite{Tegmark:1997in} and \cite{Graesser:2004ng}.}. In the context of the AHDK friendly landscape applied to realistic inflationary models 
we are able to study the probability distributions for $\epsilon, \eta$, $N$ and the scale of inflation $V_{end}$. 

In particular we find that the environmental curvature bound in the context of the AHDK friendly landscape implies a simple statistical solution to the classic $\eta$-problem of inflation \cite{Dine:1983ys,Coughlan:1984yk,Bertolami:1987xb,Dine:1995uk}, and predicts, at least for a subclass
of inflationary models favored by experiment, that the spectral index, $n_s$, of cosmological perturbations 
should deviate from one by an amount close to the current bounds.

If curvature is observed in future experiments then some of our predictions notably sharpen.
Since in future experiments the precision of tests of flatness should be improved to
$|\Omega_{\rm total}-1|_{\rm future}<10^{-3}$, the corresponding limit on the number of e-folds assuming
no curvature is seen becomes
\begin{equation}\label{NOmega}
N_{\rm future}>62+\log\left(\frac{|\Omega_{\rm total}-1|_0}{|\Omega_{\rm total}-1|_{\rm future}}\right)\approx 63.5.
\end{equation}
This number might be pushed up to $N>64.5$ in the far future.  Thus for some
of our conditional probabilities we optimistically take
the range $62< N < 64.5$ as the region in e-fold space where curvature could be seen.
We shall also discuss bounds imposed by measurements of the spectral index. Present observational
bounds on $n_s$ are discussed in \cite{Spergel:2006hy}, we will take $n_s=0.95\pm0.02$; this then
translates into bounds on the statistical distributions of slow-roll parameters $\eta$ and $\epsilon$ as
discussed below.

An important point to note is that the Coleman-deLuccia tunneling event from which, accordingly to the landscape, our universe originates, produces an open FRW universe with formally infinite volume. For this reason there is no natural measure proportional to the volume patch of the inflating universe. Thus, following \cite{FKMS}, we do not include such volume dependent factors on our measure. For a discussion of inflation in the lanscape including such volume factors, please refer to \cite{Feldstein:2005bm}. Moreover, we shall say that inflation in the landscape was previously discussed by \cite{Tegmark:2004qd} which, however, beside considering single-field instead of hybrid inflationary models,  worked in a general region of the landscape instead of the predictive 'friendly landscape' considered here.

In what follows we often chose units where the reduced Planck mass is set equal
to unity, $M_P=(8\pi G)^{-1/2}=1$.

\section{Inflation in the Landscape}

\subsection{Review of Hybrid Inflation}

Hybrid Inflation models \cite{Linde:1990gz,Linde:1991km,Lyth:1996kt,Bastero-Gil:1997vn} provide not only a flat enough potential for slow-roll inflation to
take place, but also a mechanism for it to rapidly end. This role is played by an auxiliary field
$\psi$ which, due to its interaction with the slow-rolling  field $\phi$, acquires an vacuum expectation
value only when the latter falls below a critical value $\phi_c$ and is then responsible for the
energy density of the universe after inflation.
In this model the potential for the inflaton $\phi$ and the auxiliary field $\psi$ is of the form
\begin{equation}\label{V}
V=\pm\frac{1}{2}m^2\phi^2+\frac{1}{4}\lambda(\psi^2\mp M^2)^2\pm\frac{1}{2}\lambda'\psi^2\phi^2,
\end{equation}
where the upper sign corresponds to Hybrid Inflation and the lower sign corresponds to
Inverted Hybrid Inflation.  The field $\phi$ starts at $\phi_1$ along the $\psi=0$ direction
and rolls down until it reaches the critical point $\phi_c$, where it leaves the $\psi=0$ path
and acquires an expectation value along the $\psi$ direction
\begin{equation}
\frac{\partial^2V}{\partial\psi^2}\Big\rvert_{\psi=0}=0 \quad \rightarrow \quad
\phi_{end}\approx\phi_c=\sqrt{\frac{\lambda}{\lambda'}}M.
\end{equation}
In hybrid inflation the field rolls towards the origin $\phi_c<\phi_1<\phi_{max}\approx 1$,
while in inverted hybrid inflation the inflaton rolls away from the origin $0<\phi_1<\phi_c$.
Assuming that, during inflation, the potential doesn't vary much compared to the scale of
inflation, the number of e-foldings is
\begin{equation}\label{N}
N(\phi_1,\phi_c)=\int_{\phi_c}^{\phi_1}\frac{V}{V'}d\phi=\pm\frac{\lambda}{4}
\frac{M^4}{m^2}\log\left(\frac{\phi_1}{\phi_c}\right).
\end{equation}
As displayed in Eq.(\ref{Nobsbound}) this number of e-folds must exceed $N_o \approx 62$ for inflation that
occurs near the scale $M_G\simeq 10^{16}${\rm GeV}.
However, if inflation took place at a lower scale, $V_{end}=\lambda M^4/4\ll M_G^4$,
the universe needs a shorter final epoch of inflation to achieve
the flatness we observe today.  This means that the number of e-foldings
needed will be smaller 
\begin{equation}\label{NVend}
N=\pm\frac{\lambda}{4}\frac{M^4}{m^2}\log\left(\sqrt{\frac{\lambda'}{\lambda}}
\frac{\phi_1}{M}\right)-\log\left(\frac{M_G}{V_{end}^{1/4}}\right) .
\end{equation}
We will discuss the implications of this feature below.  We will also need the amplitude
of cosmological perturbations which takes the form
\begin{equation}\label{deltaH}
\delta_H=\frac{\lambda^{3/2}}{40\sqrt{3}M_p^3\pi} \frac{M^6}{m^2\phi} .
\end{equation}
For future use we define the constant $\beta \equiv \lambda^{3/2}/40 \sqrt{3}M_p^3 \pi$.
We can assume that the cosmological scales of interest now leave the horizon approximately
at the beginning of inflation, but no more than $N_o\approx62$ e-foldings before its end,
when the value of the inflaton field was 
\begin{equation}\label{phiH}
\phi_{H}=\left\{
\begin{array}{ll}
M e^{\pm4 N_o m^2/M^4 }&\textrm{for}\quad N>N_o\\
\phi_1&\textrm{for}\quad N<N_o
\end{array}
\right.
\end{equation}
Although the case $N<N_o$ is of no practical interest, since present bounds on curvature
have already excluded this region, it is worth studying it in order to understand the
origin of these bounds in terms of probabilities. 

\begin{figure}[htbp]
\begin{center}
\includegraphics[ width=0.5\textwidth]{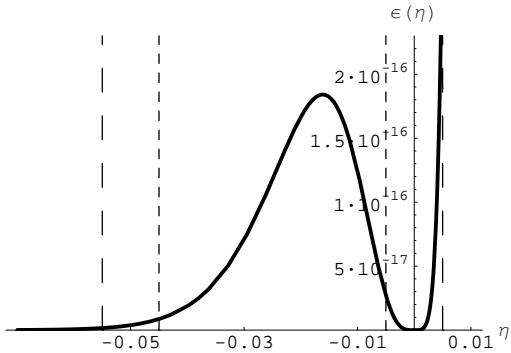}
\caption{The thick line shows the relation $\epsilon(\eta)$ for hybrid inflation models:
inverted hybrid inflation for negative $\eta$ and standard hybrid inflation for positive $\eta$.
The dashed lines represent the bounds of the observational contours at $2\sigma$ and $3\sigma$
for $n_s$ respectively, as given by \cite{Spergel:2006hy}. Constraints on standard hybrid inflation are much stronger than for inverted hybrid inflation.}
\label{EtaEpsilon}
\end{center}
\end{figure}

The scalar spectral index $n_s$ and its logarithmic variation can be written as \cite{Liddle:1992wi} 
\begin{eqnarray}
n_s&=&1-6\epsilon+2\eta\approx 1+2\eta\nonumber\\
\frac{dn_s}{d\log{k}}&=&16\epsilon\eta-24\epsilon^2-2\xi
\end{eqnarray}
where the approximation applies to all models with a high inflationary scale; it will be
proven correct in the next section. The slow-roll parameters $\epsilon$ and $\eta$ are given by
\begin{eqnarray}
\eta &=& M_{p}^2\frac{V''}{V}\approx\pm\frac{4M_P^2}{\lambda}\frac{m^2}{ M^4}\nonumber\\ 
\epsilon &=& \frac{M_{p}^2}{2}\left(\frac{V'}{V}\right)^2
\approx\frac{8M_p^2}{\lambda^2}\frac{m^4\phi^2}{M^8},
\label{etaepsilon}
\end{eqnarray}
($\xi$ vanishes in this model) and the ratio of tensor to scalar density perturbations is given directly in terms of $\epsilon$,
\begin{equation}
r=12.4\,\epsilon.
\end{equation}
In this model the shape of the inflationary potential is determined by a small number of
parameters and constraints on $\delta_H$ and $N$ translate into a relation between $\epsilon$ and $\eta$,
\begin{equation}\label{epsilon(eta)}
\epsilon(\eta)=\frac{e^{4 N_o \eta} \delta_H^2 \eta^4}{32\beta^2},
\end{equation}
as shown in Figure~\ref{EtaEpsilon}.
Since we are interested in the amplitude of cosmological perturbations and the value of the
slow-roll parameters at the present epoch, we shall evaluate both $\epsilon$ and $\delta_H$ at $\phi_H$.
In what follows, we will make the natural choice for the dimensionless parameters
$\lambda\sim\lambda'\sim1$, implying, $\beta\approx 218^{-1}$.

\subsection{Distributions and Signals in Inverted Hybrid Inflation}

In inverted hybrid inflation  \cite{Lyth:1996kt,Bastero-Gil:1997vn} the field rolls away from the
origin towards its critical value $\phi_c$ along a potential slope with negative curvature. This negative
curvature is responsible for an infrared spectrum of cosmological perturbations, $n_s<1$, as favored
by the latest WMAP data \cite{Spergel:2006hy}.

As discussed in the introduction, in the framework of the AHDK `friendly landscape' \cite{ADK}, unprotected
dimensionful quantities are scanned over, while dimensionless quantities are essentially fixed. Therefore,
in the string theory ensemble of inflation theories described by the
potential Eq.(\ref{V}), one should scan over $\phi_1$ (the value of the inflaton at the beginning of inflation),
$M$ and $m$. While no particular bounds apply on $0<m<M_{p}=1$ and $0<M<1$, the requirement
$N>0$ imposes $0<\phi_{1}<M$ for the starting field value.\footnote{In principle $M$ is bounded from
below by the requirement that the success of Big Bang Nucleosynthesis is not compromised, and more model-dependently,
that at least low-scale electroweak baryogenesis can occur, implying roughly $M>1$ TeV.  These bounds are too
small to have any influence on the present discussion.}

\subsubsection{Combined Distribution of $N$ and $\eta$}\label{SectionPNepsilon}

We begin our discussion on distributions of inflationary parameters by studying the
combined distribution of the slow-roll parameter $\eta$ and the number of e-foldings $N$,
from which we can deduce all the relevant distributions:
\begin{equation}
\begin{array}{rl}
P(N,\epsilon)\equiv \int_{0}^1dm\int_{0}^1dM & \, \, \int_{0}^Md\phi_1 ~ \delta\left(N-\frac{M^4}{4m^2}\ln
\left(\frac{M}{\phi_1}\right)\right)\qquad\qquad\qquad\\
&\qquad\qquad\qquad\delta\left(\delta_H-\frac{\beta M^6}{m^2\phi_H}\right)\delta\left(\eta+\frac{4 m^2}{M^4}\right).
\end{array}
\label{P(Ne)}
\end{equation}
We first study the case $N>N_o$, the case $N<N_o$ can be deducted from the
previous one by setting $N_o=N$ (in this region the slow roll parameters are evaluated at
$N_o=N$, i.e. at the beginning of inflation). In this case $\phi_H=e^{-4 N_o m^2/M^4 }$.
The space of solutions consists of a single point in the 3-dimensional space (${m,M,\phi_1}$),
and the delta-functions can therefore be solved without ambiguity, resulting in the Jacobian
\begin{equation}
|J|^{-1}  = \frac{m^3\phi_1 e^{-4\frac{m^2N_o}{M^4}}}{2M^4\beta}
\end{equation}
evaluated at the unique solution of the $\delta$-functions
\begin{equation}\label{SolutionNepsilon}
m(N,\eta)=\frac{(-\eta)^{5/2}e^{2N_o\eta}\delta_H^2}{32\beta^2},~~M(N,\eta)=-\frac{\eta e^{N_o\eta}\delta_H}{4\beta},
~~\phi_1(N,\eta)=-\frac{\eta e^{(N_o+N)\eta}\delta_H}{4\beta} .
\end{equation}
This gives the combined distribution (non normalized)
\begin{equation}\label{PNeta}
P(N,\eta)=\frac{\delta_H^3 e^{\eta(4N_o+N)}(-\eta)^{9/2}}{2^{10}\beta^4},
\end{equation}
where, for $N<N_o$ we have to set $N_o=N$.  

At this point we can check the consistency of the assumption made in 
Eq.(\ref{etaepsilon}), that the scale of inflation is much bigger than its change during
slow-roll inflation: one might wonder if this is still the case when the variables $M$ and
$m$ are scanned over. This approximation corresponds to $\frac{1}{2}m^2\phi^2\ll \frac{M^4}{4}$, or 
\begin{equation}
\frac{2m^2\phi^2}{M^4}=-\frac{\eta^3 e^{2(N_o+N)\eta}\delta_H^2}{32\beta^2}\ll1
\end{equation}
which, due to $\eta$ being negative, is exponentially small, and thus the assumption is consistent.

If, as suggested by Denef and Douglas \cite{Douglas2},
the fundamental quantities to be scanned over,
are $dm^2=2mdm$ and $dM^2=2MdM$ and not $dm$ and $dM$ as assumed here, then the distribution
Eq.(\ref{PNeta}) changes.  However, the present results are only marginally sensitive to such
a change in scanned parameters: the distribution $P(N,\eta)$ becomes proportional to
\begin{equation}
P_{M^2m^2}(N,\eta)=\frac{\delta_H^6 e^{\eta(7N_o+N)}(-\eta)^{8}}{32768 \beta^7}
\end{equation}
but the domain of definition is unaltered.  Although this would result
in slightly steeper distributions the main features remain unchanged and our qualitative
results will be similar.

\begin{figure}[htbp]
\begin{center}
\includegraphics[ width=0.5\textwidth]{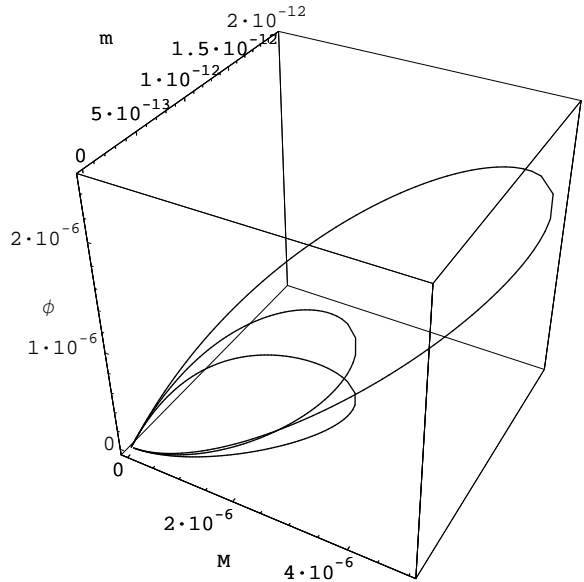}
\caption{The domain of integration in $(M,m,\phi_1)$ space as constrained by requiring the
right amplitude of cosmological perturbations $\delta_H\approx10^{-5}$ for $N=40$ (longer curve), $N=62$ and $N=100$.}
\label{figInvertedDomain}
\end{center}
\end{figure}

Generally, one must make sure that the definition domain of this distribution is
compatible with the original integration domain in $(M, m, \phi_1) <1$ and the bounds
imposed by the delta functions. As shown in Figure~\ref{figInvertedDomain}, however, for
a given $N$ and $\delta_{H}$, the delta functions naturally constrain $(M,m,\phi_1)$ well
inside their external bounds. This is an interesting feature for this kind of computation,
where parameters are left free to vary within a given range, because it reduces the
amount of input information to a minimum.

As mentioned in Eq.(\ref{NVend}), if inflation took place at a lower scale
$M^4/4\ll M_G^4$, then the number of e-foldings needed to solve the flatness problem is
reduced $N\rightarrow N-\log(\sqrt{2}M_G/M)$. Since $M$ is being
integrated over, this contribution could become significant -- its net effect is a shift
in the $(N,\eta)$ plane
\begin{equation}\label{Shift}
4N_o+N\rightarrow 4N_o+N-\frac{6+\eta}{1+\eta}\log\left(\frac{4\sqrt{2}M_G\beta
e^{-N_o \eta}}{\delta_H(-\eta)}\right)-\frac{5}{2}\log(2).
\end{equation}
In what follows we will mainly discuss results following directly from Eq.(\ref{PNeta})
and analyze Eq.(\ref{Shift}) only numerically.

\subsubsection{Distribution of $N$}

\begin{figure}[htbp]
\begin{center}
\includegraphics[ width=0.5\textwidth]{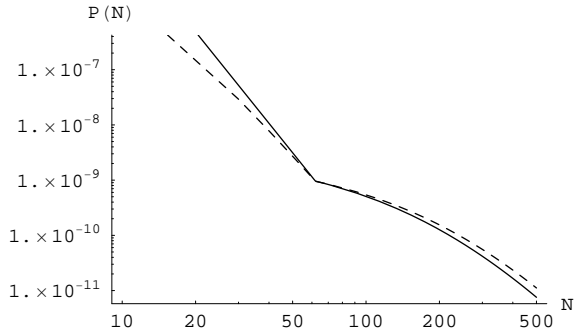}
\caption{The distribution of N in the inverted hybrid inflation model. The dashed curve
includes the correction from $V_{end}\ll M_G^4$.}
\label{InvertedPofN}
\end{center}
\end{figure}

To obtain the absolute distribution of the number of e-foldings, regardless of the value of
$\eta$, we integrate Eq.(\ref{PNeta}) over the proper range $\eta\in[\eta_{min},0]$
(recall $\eta_{min}<0$). The non-normalized distribution is given by
\begin{equation}\label{PN}
P(N)=\frac{\delta_H^3}{2^{10}\beta^4}\frac{1}{(4N_o+N)^{11/2}}\int_0^{(4N_o+N)|\eta_{min}|}
t^{9/2}e^t dt\approx\frac{945\sqrt{\pi}}{2^{14}\beta^{4}}\frac{1}{(4N_{o}+N)^{11/2}},
\end{equation}
which, for $N<62$ decreases  as $(N)^{-11/2}$ and for $N>62$ as $(N+248)^{-11/2}$, as shown
in Figure~\ref{InvertedPofN}. This gives a $97\%$ probability that the present observational bound $N>N_o$ is satisfied, given the anthropic bound (\ref{Nobsbound}). However, the likelihood of finding curvature in future experiments, $62<N<64.5$, reduces to less than $3\%$, thus disfavoring the idea that 'just so' inflation might be favored.\footnote{This is in contrast with the work of Freivogel {\it et al}, where the distribution of the number of e-foldings was proportional to $N^{-4}$. In FKMRS, however,  beside a different weight of integration, the constraint on $\delta_H$ is imposed at the beginning of inflation, independently of the duration of inflation.}
It is worth noting that the result is exponentially insensitive to the lower bound of integration $\eta_{min}$, i.e. it doesn't influence the distribution whether we impose the
biased observational bound $\eta_{min}=(n_{s}-1)/2$ or let the integration run
all the way down to $\eta_{min}=-1$. 

In this case, the shift due to Eq.(\ref{Shift}) doesn't have an important influence on the distribution,
its effect is also plotted on Figure~\ref{InvertedPofN}.

\subsubsection{Distribution of $\eta$}

\begin{figure}[htbp]
\begin{center}
\includegraphics[ width=0.5\textwidth]{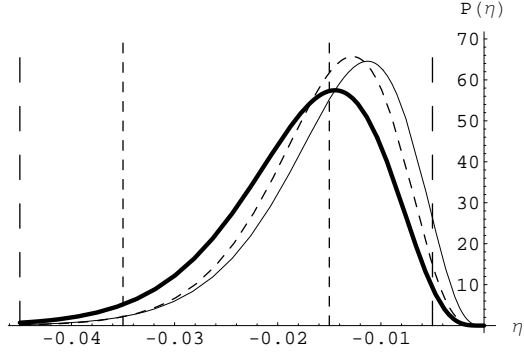}
\caption{The distribution of $\eta$ for the different cases, $N\in [62,\infty]$ (thin curve), $N\in [62,64.5]$
(thick curve) and including corrections from $V_{end}\ll M_G^4$ (dashed curve). The vertical dashed
lines represent the observational $1\sigma$ and $2\sigma$ bounds respectively, as given
by \cite{Spergel:2006hy}}
\label{InvertedPeta}
\end{center}
\end{figure}

We now would like to calculate the distribution of $\eta$, given observational bounds 
on $N$. This implies integrating $P(N,\eta)$, Eq.(\ref{PNeta}), over $N\in[N_{min},N_{Max}]$.
If curvature is in the present allowable range, $N\in[N_o=62,\infty]$, then the distribution of $\eta$ becomes
\begin{equation}\label{InvertedPetaEq}
P(\eta)_{N\in[62,\infty]}=\frac{\delta_H^3 e^{\eta(5N_o)}(-\eta)^{7/2}}{2^{10}\beta^4},
\end{equation}
which, as shown in Figure~\ref{InvertedPeta}, is strongly peaked around its average
value at $\bar{\eta}\approx -0.015$.  This is an important result of our analysis, in fact it states that the probability that $\eta$ is within $2\sigma$ of its present observational bounds $\eta\in[-0.045,-0.005]$ is as big as $96\%$, compared to $0.1\%$ to find smaller $\eta<-0.045$ and $3.9\%$ for $\eta\in[-0.005,0]$, in an unobservable small range for deviations of $n_s$ from unity . Thus, the requirement of structure formation (\ref{Nebound}) provides a solution to the classic $\eta$-problem in terms of probability distributions.

Unless curvature happens to be undetectably small, corresponding to $N\gg62$, then
the distribution of $\eta$ is quite insensitive to its
upper detection $N$ bound. For example, if curvature is found in future experiments
between $N\in[62,64.5]$, the distribution becomes proportional
to $P(\eta)_{N\in[62,64.5]}\propto e^{\eta(5N_o)}(-\eta)^{7/2}(1-e^{(64.5-62)\eta})$. Its average value would be $\eta\approx-0.017$ and the probabilities of finding $\eta\in[-0.045,-0.005]$, $\eta<-0.045$ or  $\eta\in[-0.005,0]$ are $98.6\%$, $0.4\%$ and $1\%$ respectively. Both functions are shown in Figure~$\ref{InvertedPeta}$
and compared with the numerical computation of $P(\eta)$ including the correction Eq.(\ref{Shift}).

\subsubsection{Distribution of $\epsilon$ and Inflationary Scales}
\label{SectionEpsilonInverted}
The distribution of $\epsilon$ can be obtained similarly to the one with $\eta$: evaluating
the Jacobian along the solution of the equations for $N$, $\delta_H$ and $\epsilon$.
Alternatively, one can derive the $\epsilon$-distribution from the $\eta$-distribution.
This is possible, thanks to the relation Eq.(\ref{epsilon(eta)}) between $\epsilon$ and $\eta$,
\begin{equation}\label{PetaPepsilon}
P(\epsilon)=P(\eta(\epsilon))\frac{d}{d\epsilon}\eta(\epsilon).
\end{equation}
The function $\eta(\epsilon)$, the inverse of Eq.(\ref{epsilon(eta)}), has two different
branches of solutions for negative $\eta$,
\begin{equation}
\eta(\epsilon)=\frac{1}{N_{o}}w\left(0,-\frac{2^{5/4}\sqrt{\beta} N_{o}
\epsilon^{1/4}}{\sqrt{\delta_H}}\right)~~\textrm{and}~~\eta(\epsilon)=
\frac{1}{N_{o}}w\left(-1,-\frac{2^{5/4}\sqrt{\beta} N_{o} \epsilon^{1/4}}{\sqrt{\delta_H}}\right)
\end{equation}
where $w(0,x)=w(x)$ and $w(-1,x)$ are the Lambert w-function and the generalized Lambert
$w$-function respectively (different branches of the inverse of $xe^x$). The distribution of
$\epsilon$ for $N\in[N_{min},N_{Max}]$ becomes
\begin{equation}\label{PepsilonStandard}
P(\epsilon)=Q_{N_{min}}^{(0)}(\epsilon)-Q_{N_{min}}^{(-1)}(\epsilon)-Q_{N_{Max}}^{(0)}(\epsilon)+Q_{N_{Max}}^{(-1)}(\epsilon)
\end{equation}
where
\begin{equation}
Q_{N}^{(k)}(\epsilon)=\frac{\delta_{H}\sqrt{w\left(k,-\frac{2^{5/4}N_o\epsilon^{1/4}}{\sqrt{\delta_H/\beta}}\right)}}
{2^7\sqrt{N_o}\beta^2\left(1+w\left(k,-\frac{2^{5/4}N_o\epsilon^{1/4}}{\sqrt{\delta_H/\beta}}\right)\right)}
\left(-\frac{\frac{2^{5/4}N_o\epsilon^{1/4}}{\sqrt{\delta_H/\beta}}}{w\left(k,-\frac{2^{5/4}N_o
\epsilon^{1/4}}{\sqrt{\delta_H/\beta}}\right)}\right)^{\frac{N}{N_o}}.
\end{equation}
The term exponentiated to the $N$-th power is smaller than one, and therefore vanishes in the $N\rightarrow\infty$ case.  We can, again, discuss the cases where curvature is in the present observable bounds or it will be detected in the near future.  In both cases, the distribution of $\epsilon$ increases with $\epsilon$ and has a cut-off when the argument of the $w$-function becomes $1/e$, there the
function quickly approaches  -1 and the denominator vanishes; this happens for
\begin{equation}\label{Invertedepsilonmax}
\epsilon_{Max}=\frac{\delta_{H^2}}{32 e^4N_{o}^4\beta^2}\approx 1.8\times 10^{-16}.
\end{equation}
This natural constraint on $\epsilon$ justifies the choice  $\eta\gg\epsilon$ in previous sections and suggests that, though the upper bound (\ref{Invertedepsilonmax}) is preferred by the distribution, tensor perturbations will never be detected. 

\begin{figure}[htbp]
\begin{center}
\includegraphics[ width=0.5\textwidth]{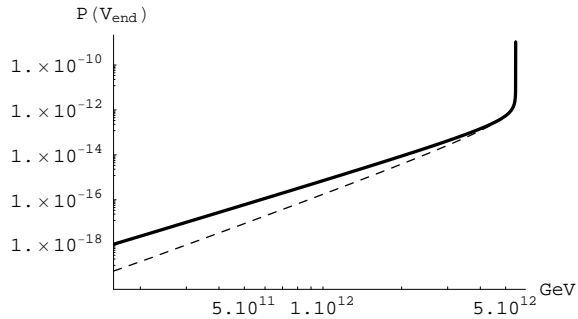}
\caption{This picture shows the distribution of inflationary scales $V_{end}$ plotted in logarithmic scale for inverted hybrid inflation: the dashed curve represents $N\in [62,63.5]$, while the thick curve $N\in [62,\infty]$. Intermediate scales of inflation are strongly preferred in this model.} 
\label{InvertedPepsilon}
\end{center}
\end{figure}

In this model, we can use  (\ref{epsilon(eta)}) and (\ref{SolutionNepsilon}) to express the scale of inflation $V_{end}^{1/4}=M/\sqrt{2}$ in terms of $\epsilon$,
\begin{equation}\label{Vendepsilon}
V_{end}=\frac{\epsilon^{1/4}\sqrt{\delta_H}}{2^{5/4}\sqrt{\beta}}
\end{equation}
and easily express the distribution of inflationary scales in terms of the ditribution of $\epsilon$: $P(V_{end})=P_{\epsilon}(\epsilon(V_{end}))\frac{d}{dV_{end}}\epsilon(V_{end})$, where $\epsilon(V_{end})$ is the inverse of (\ref{Vendepsilon}) and $P_{\epsilon}(\epsilon)$ is given by (\ref{PepsilonStandard}).  Such distribution is plotted in Figure \ref{InvertedPepsilon} and shows that the upper bound 
\begin{equation}
V_{end}\approx5.4 \times10^{12}{\rm GeV},
\end{equation}
corresponding to an intermediate scale of inflation, is strongly preferred: for example, the probability that the scale of inflation be between $V_{end}\in[4.4\times10^{13}{\rm GeV},5.4\times10^{13}{\rm GeV}]$ is about $70\%$.

\subsubsection{Distribution of $\Omega$ and $n_s$}\label{SectionnOmega}

\begin{figure}[htbp]
\begin{center}
\includegraphics[ width=0.5\textwidth]{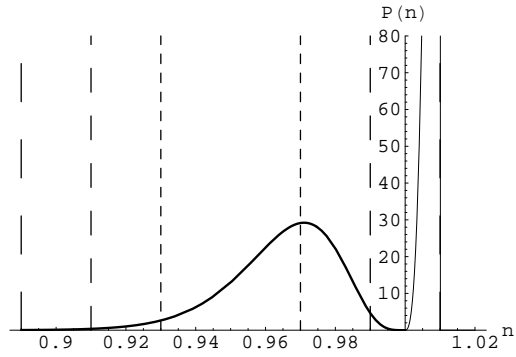}
\caption{The distribution $P(n)$ of spectral indexes in inverted hybrid inflation (thick line, negative $\eta$) and standard hybrid inflation (continuous line, exponentially growing, positive $\eta$).  The vertical lines represent present observational bounds, at 1, 2 and 3 $\sigma$ respectively.}
\label{InvertedPn}
\end{center}
\end{figure}

\begin{figure}[htbp]
\begin{center}
\includegraphics[ width=0.5\textwidth]{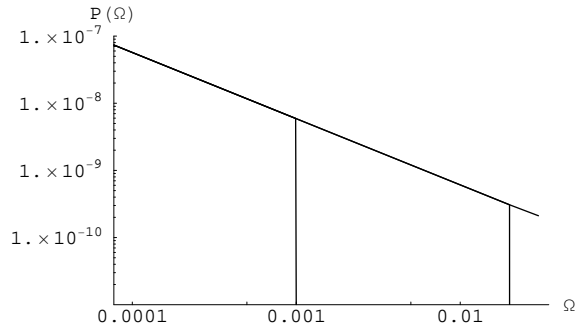}
\caption{The distribution of $\Omega$ is dominated by its exponential dependence on the number of
e-foldings $N$, resulting mostly in a $P(\Omega)\sim\Omega^{-1}$ distribution. The vertical lines
show both the present observational bound at $\Omega<0.02$ and the precision of future
experiments, $\Omega<0.001$.}
\label{InvertedPOmega}
\end{center}
\end{figure}

It is interesting to express these distributions also in terms of directly observable quantities, such as the
index of density perturbations, $n_{s}$ and the curvature $\Omega$. While the former can be expressed by making
the simple variable change $\eta=(n_{s}-1)/2$ in Eq.(\ref{InvertedPetaEq}), the latter is given by 
\begin{equation}
P(\Omega)=P_N(N(\Omega))\frac{d}{d\Omega}N(\Omega)=\frac{1}{\Omega}P_{N}(62+\log\left(\frac{0.02}{|\Omega-1|}\right)),
\end{equation}
where we have used Eq.(\ref{NOmega}) and $P_N(N)$ is given by Eq.(\ref{PN}). Due to the exponential dependence of
$\Omega$ on $N$, the distribution of $\Omega$ parameters will always be dominated by the $\Omega^{-1}$ factor,
resulting from the change of variables, unless $P(N)$ is exponentially falling. These distributions are plotted
in Figures \ref{InvertedPn} and \ref{InvertedPOmega} respectively.

\subsection{Distributions and Signals in Standard Hybrid Inflation}

It is interesting to compare different models of inflation and investigate the model-dependence
of the distributions considered above. In what follows we shall apply the methods of the previous section to standard hybrid inflation.
 
\subsubsection{Combined Distribution of $N$ and $\eta$}

\begin{figure}[htbp]
\begin{center}
\includegraphics[ width=0.5\textwidth]{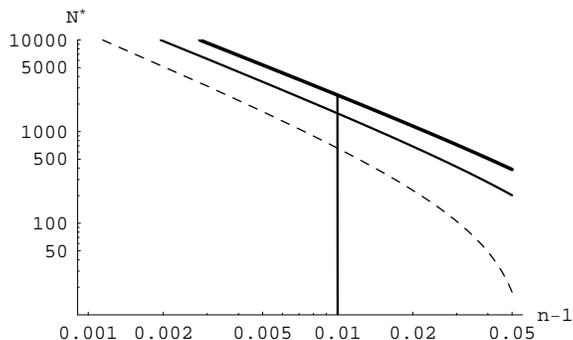}
\caption{This figure shows $N^*$ as a function of the spectral index $n_s$. The thick
line is for $\phi_{max}=1$, while the thin one corresponds to $\phi_{max}=0.01$ and
the dashed one to $\phi_{max}=0.0001$. The vertical line represent the upper bound of the $3\sigma$ contour for positive $n_s$.}
\label{figNStar}
\end{center}
\end{figure}

Similarly to the inverted hybrid inflation case, we need
\begin{equation}\label{P(Ne)hybrid}
\begin{array}{lr}
P(N,\eta)=&\int_{0}^1dm\int_{0}^1dM\int_{M}^1d\phi_1\delta\left(N-\frac{M^4}{4m^2}
\log\left(\frac{\phi_1}{M}\right)\right)
\delta\left(\delta_H-\beta\frac{M^6}{m^2\phi_H}\right)\qquad\qquad\\
&\times\delta\left(\eta-\frac{m^2}{ M^4}\right)\Theta\left(n_{Max}-\left(1+2\eta\right)\right),
\end{array}
\end{equation}
where now, due to the stricter observational bounds on $n_{s}>1$, we need to constrain the
definition domain via  the $\Theta$ function to $n_s\in[0,n_{Max}]$. The Jacobian resulting
from the delta-functions is, of course, identical to Eq.(\ref{PNeta}), except form the change
$(-\eta)\rightarrow\eta$; the combined distribution of $N$ and $\eta$ becomes
\begin{equation}\label{PNeta2}
P(N,\eta)=\frac{\delta_H^3 e^{\eta(4N_o+N)}\eta^{9/2}}{2^{10}\beta^4}.
\end{equation}
However, the domain of integration is now completely different. In first instance,
since $\eta$ is now positive, $m$, $M$ and $\phi_{1}$ in Eq.(\ref{SolutionNepsilon})
are now exponentially growing functions of $\eta$, therefore the domain of integration
is no longer self-constrained, but needs to be bound from above by $\phi<\phi_{Max}=1$,
which is the first of the scanned parameters to reach its maximum value along the constrained solution of Eq.(\ref{deltaH}) and Eq.(\ref{N}).
This translates into a bound on $\eta$
\begin{equation}\label{etamaxphi}
\eta<\eta_{\phi}(\phi_{Max},N)=\frac{w(4(N_o+N)\phi_{Max}\beta/\delta_H)}{N+N_o}.
\end{equation}
On the other hand, we also have the $\Theta$-function to take into account, this reduces
the definition domain to
\begin{equation}\label{etamaxn}
\eta<\eta_n(n_{Max})=\frac{n_{Max}-1}{2}.
\end{equation}
In general the latter is smaller than the former, however, since the first bound decreases
with increasing $N$, for a specific value $N^*$ they will have the same value, and for
$N>N^*$ Eq.(\ref{etamaxphi}) will be the stronger constraint. $N^*$ is given by
\begin{equation}
N^*=\frac{2}{n_s-1}\log\left(\frac{8 \beta \phi_{Max}}{\delta_H(n_s-1)}\right)-N_o,
\end{equation}
and is plotted in Figure~\ref{figNStar}. It slowly decreases with decreasing
range of $\phi\in[0,\phi_{Max}]$, but increases as fast as $1/(n_{Max}-1)$ when the upper bounds
on $n_s$ approach unity. Thus, since present observational constraints are very strict on positive $n_s$, $N^*$ becomes a very large number.

\subsubsection{Distribution of $N$}\label{SectionN}

\begin{figure}[htbp]
\begin{center}
\includegraphics[ width=0.5\textwidth]{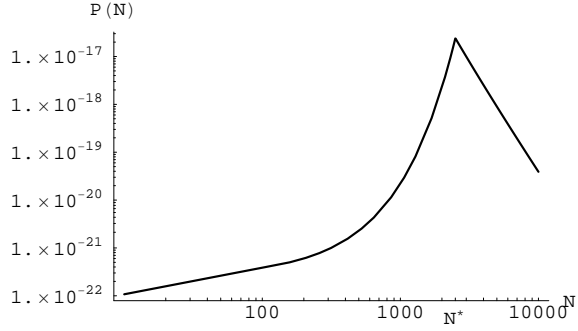}
\caption{The distribution of N for positive $n_s$ within $3\sigma$ of its observed value and for $\phi_{Max}=1$.}
\label{DistN}
\end{center}
\end{figure}

To obtain the absolute distribution of the number of e-foldings, we need to integrate Eq.(\ref{PNeta2})
over the whole range of definition in $\eta$. Since both the integrand and the limits of
integration are piecewise functions of $N$, this requires some care. For $N<N^*$ the bound set by
the constraints on the spectral index Eq.(\ref{etamaxn}) is always smaller than the bound coming
from the integration domain Eq.(\ref{etamaxphi}), therefore, for these values of $N$ the
integration over $\eta$ is bound by Eq.(\ref{etamaxn}) and the integrand is given by Eq.(\ref{PNeta2}),
where, for $N<N_o=62$ we set $N_{o}=N$.
\begin{equation}
P(N)_{N<N^*}=\frac{\delta_H^3}{2^{10}\beta^4}\frac{1}{(4N_o+N)^{11/2}}\int_0^{(4N_o+N)\frac{n_s-1}{2}}t^{9/2}e^t dt
\end{equation}
This distribution is exponentially peaked towards big values of N. 
Moreover, when $N$ increases above $N^*$ the definition domain of $\eta$ becomes smaller
and the integral runs only up to Eq.(\ref{etamaxphi}), we obtain
\begin{equation}
\begin{array}{lll}
P(N)_{N^*<N}&=&\frac{\delta_H^3}{2^{10}\beta^4}\frac{1}{(4N_o+N)^{11/2}}
\int_0^{\frac{4N_o+N}{N_o+N}w(4(N_o+N)\beta/\delta_H)}t^{9/2}e^t dt\\
&\approx &\frac{\sqrt{\beta}}{2\delta_H^{3/2}(4N_o+N)}\left(
\frac{w(4(N_o+N)\beta/\delta_H)}{4(N_o+N)\beta/\delta_H}\right)^{7/2-3\frac{N_o}{N_o+N}}
\end{array}
\end{equation}
where we used $\int_0^{t_{max}} t^{9/2}e^t dt \approx t_{max}^{9/2}e^{t_{max}}$. The resulting
distributions decreases very fast, due to both the negative powers of $N$ and the exponential.
Altogether this shows that, in this model, the number of e-foldings wants to sit exponentially
close to $N^*$. The distribution is plotted in Figure \ref{DistN} and shows the sharp peak of
probability around $N^*$, way beyond any reasonably observable value.

Again, we can discuss the corrections due to $V_{end}\ll M_G^4$. A detailed calculation shows that
the net effect of this correction is a shift in $P(N)$ as
\begin{equation}
N^*\rightarrow N^*-\log\left(\frac{8\sqrt{2}M_G \beta}{\delta_H(n_s-1)e^{(n_s-1)N_o/2}}\right),
\end{equation}
which, however, is always smaller than order ten and doesn't have important consequences on the distributions around $N^*\sim1000$.

\subsubsection{Distribution of $\eta$}

The distribution of $\eta$ depends on the range where curvature is expected. We are interested
in the region $N>N_o$ where the probability of finding $\eta$ given
$N\in [N_{min},N_{Max}]$ takes the form
\begin{equation}\label{Peta}
P(\eta)\propto \left.\frac{\delta_H^2}{8\beta^2}\eta^{7/2}
e^{(N+4N_o)\eta}\right|_{N=N_{min}}^{N=N_{Max}}.
\end{equation}
For curvature in the present observable range, $N\in[62,\infty]$, we shall set $N_{min}=62$
in Eq.(\ref{Peta}), while its upper bound will be given by the inverse of Eq.(\ref{etamaxphi})
as a function of $N$ (since $N=\infty$ lies outside the domain of definition of the $P(N,\eta)$ function). The result is
\begin{equation}
P(\eta)_{N\in[62,\infty]}\propto\frac{e^{3 N_o \eta}\delta_H \eta^{5/2}\phi_{Max}}{2\beta}
\left(1-\frac{e^{2N_o\eta}\delta_H\eta^{5/2}}{4\beta\phi_{Max}}\right)
\end{equation}
which is exponentially growing for $\eta$ small enough, which is always true, since
$\eta<\frac{n_s-1}{2}<0.005$.

If curvature turns out to be observable in the near future,  $N\in[62,64.5]$, we can
approximate $e^{(N_{Max}-N_{Min})\eta}\approx1+(N_{Max}-N_{Min})\eta$ and obtain from
Eq.(\ref{Peta})
\begin{equation}
P(\eta)_{N\in[62,63.5]}\sim\frac{e^{5 N_o \eta}\delta_H^2 \eta^{9/2}\phi_{Max}}{8\beta^2}(N_{Max}-N_{min})
\end{equation}
which also grows exponentially in up to Eq.(\ref{etamaxn}).

\begin{figure}[htbp]
\begin{center}
\includegraphics[ width=0.5\textwidth]{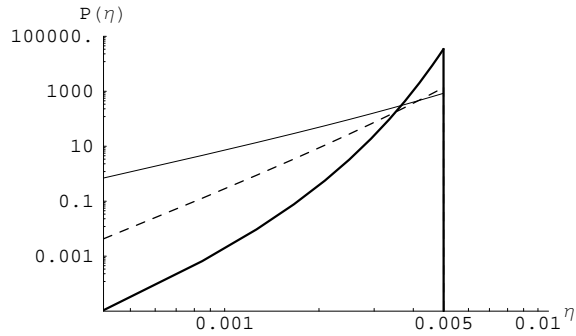}
\caption{This picture shows the distribution of $\eta$ parameters, plotted in logarithmic
scale, for the different cases, $N\in [62,\infty]$ (thin curve), $N\in [62,63.5]$ (dashed curve) and $N$ around $N^*$ (thick curve).}
\label{DistEta}
\end{center}
\end{figure}

Finally we can consider the scenario suggested by the discussion of Section \ref{SectionN},
when $N=N^*$. In this case the distribution of $\eta$ grows exponentially as
\begin{equation}
P(\eta)_{N=N^*}=\frac{\delta_H^3 e^{\eta(4N_o+N*)}\eta^{9/2}}{2^{10}\beta^4}.
\end{equation}
This distribution is shown for these three cases in Figure~\ref{DistEta},
suggesting that, independently from whether curvature is found in future experiments or not,
the distribution of $\eta$ will be peaked towards as large as possible positive values. Corrections
coming from $V_{end}\ll M_G^4$ are only logarithmic in $\eta$ and have no influence on
such behavior.

As discussed in Section~\ref{SectionnOmega}, we can interpret  this result as a
distribution for $n_s$, the result is also plotted in Figure~\ref{InvertedPn}.

\subsubsection{Distribution of $\epsilon$ and Inflationary Scales}

We can use expression Eq.(\ref{PetaPepsilon}) to deduce the distribution of $\epsilon$
from $P(\eta)$, Eq.(\ref{Peta}). This gives
\begin{equation}\label{Pepsilon}
P(\epsilon)=\frac{\sqrt{w\left(\frac{2^{5/4}N_o\epsilon^{1/4}}{\sqrt{\delta_H/\beta}}\right)}}
{\sqrt{N_o}\left(1+w\left(\frac{2^{5/4}N_o\epsilon^{1/4}}{\sqrt{\delta_H/\beta}}\right)\right)}
\left.\left(\frac{\frac{2^{5/4}N_o\epsilon^{1/4}}{\sqrt{\delta_H/\beta}}}{w\left(\frac{2^{5/4}N_o\epsilon^{1/4}}
{\sqrt{\delta_H/\beta}}\right)}\right)^{\frac{N}{N_o}}\right|_{N=N_{min}}^{N=N_{Max}}.
\end{equation}
Where the upper bound on the spectral index translates into an upper bound for $\epsilon$
\begin{equation}\label{emax}
\epsilon_n(n)\approx\frac{e^{2N_o(n_{s}-1)}(n_{s}-1)^4\delta_H^2}{512\beta^2}\approx 3\times 10^{-16},
\end{equation}
which is, again, too small to have any influence on detectable quantities, such as the ratio of tensor to scalar perturbations, $r$. If curvature is in the
present observable range, $N\in[62,\infty]$, $P(\epsilon)$ behaves approximately like 
\begin{equation}
P(\epsilon)\sim \epsilon^{-1/4},
\end{equation}
and propounds  for a vanishing $\epsilon$. On the other hand, if we detect curvature in near
future experiments, $N\in[62,63.5]$ the situation is different, we obtain a distribution
approximately proportional to 
\begin{equation}
P(\epsilon)\sim\epsilon^{1/4},
\end{equation}
therefore slightly preferring higher values of $\epsilon$. If finally, as suggested by
the previous analysis of $P(N)$, the number of e-foldings sits around $N=N^*$, then
\begin{equation}
P(\epsilon)\sim\epsilon^{\frac{N^*}{4N_o}}.
\end{equation}

\begin{figure}[htbp]
\begin{center}
\includegraphics[ width=0.5\textwidth]{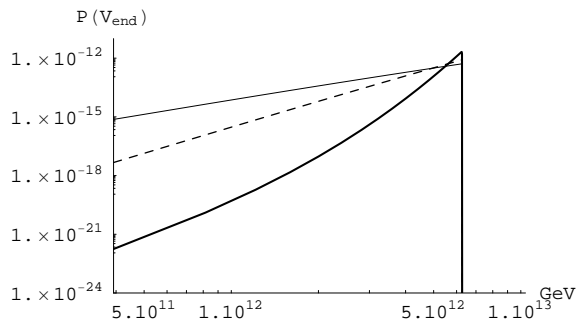}
\caption{The distribution of inflationary scales $V_{end}$ in standard hybrid inflation: the thin line represents $N\in [62,\infty]$,  the dashed curve $N\in [62,63.5]$ and $N$ around $N^*$ corresponds to the thick line.}
\label{DistEpsilon}
\end{center}
\end{figure}

Following section \ref{SectionEpsilonInverted}, we transform the distribution of $\epsilon$ into a distribution for the inflationary scale $V_{end}$. The result is a power-law growing function of $V_{end}$, as plotted in Figure~\ref{DistEpsilon}, with a cut-off, due to observational bounds on the spectral index, for scales bigger than
\begin{equation}
V_{end}\approx6.2\times 10^{12} {\rm GeV}.
\end{equation}
Although the distribution of inflationary scales is growing only as a power-law function, the fact that we are interested in a wide logarithmic range in $V_{end}$ means that this distribution is indeed exponentially peaked towards intermediate scales of inflation. With the present bounds on curvature, the probability for the inflationary scale to be in  $V_{end}\in[4.2\times10^{13}{\rm GeV},6.2\times10^{13}{\rm GeV}]$ is already about $70\%$. 

\section{Conclusions}
String theory appears to predict the existence of an exponentially large number of vacuum states --
{\it the landscape} -- corresponding to different possible universes with
different values of the parameters governing the laws of nature.
Following Freivogel {\it et al} \cite{FKMS} we considered inflation in a predictive (or `friendly')
region of the landscape of string vacua, as modeled by Arkani-Hamed, Dimopoulos and Kachru \cite{ADK}. 
In such a region the dimensionful coefficients of super-renormalizable operators unprotected
by symmetries, such as the vacuum energy and scalar mass-squareds are freely scanned over,
and the objects of study are anthropically or `environmentally' conditioned probability distributions
for observables.  In this context we studied the statistical predictions of Inverted Hybrid
Inflation and for Standard Hybrid Inflation
models, where the properties of the inflaton are probabilistically distributed, and we derived
the resulting distributions of slow-roll parameters, $\epsilon$ and $\eta$, and the number
of e-folds of inflation, $N$, and thus the distributions of observables, including the
deviation from flatness $|1-\Omega|$, the spectral index of scalar cosmological perturbations $n_s$
(and its scale dependence $dn_s/d\log k$), and the ratio of tensor to scalar perturbations $r$.
In the Inverted Hybrid Inflation case, the assumption (following Weinberg's anthropic
study of the cosmological constant) that the amplitude of
cosmological perturbations be of the right size $\delta_H\approx 10^{-5}$, together with the environmental bound on the curvature, are enough to provide both an
environmental solution to the $\eta$ problem and to predict measurable deviations of the
spectral index $n_s$ from 1, well within present observational constraints (see Figure
\ref{InvertedPn}).  We find a relatively low probability ($<3\%$) of `just-so' inflation
with measurable deviations from flatness.
For Standard Hybrid Inflation, on the other hand, present observations
already constrain the domain of the distributions of slow-roll parameters to a confined
region $\eta<0.005$.  If future experiments reach high enough precision, then this
model predicts that the upper bound of the region of $\eta$ is exponentially preferred. 
The distribution of the number of e-foldings for Standard Hybrid Inflation
prefers a large number of e-foldings, beyond any possibility of a future detection of curvature.
In both cases the slow-roll parameter $\epsilon$ is far too small to
be ever detected via tensor perturbations in the CMB radiation,
$\epsilon\sim 10^{-16}$, and an intermediate scale of inflation ($\sim 10^{12}${\rm GeV})
is strongly preferred (see Figure \ref{InvertedPepsilon}).

\section{Acknowledgments}

The authors thank Arthur Hebecker, Subir Sarkar and especially Nima Arkani-Hamed and Savas Dimopoulos for
discussions. This work has been supported by PPARC Grant PP/D00036X/1, and
by the `Quest for Unification' network, MRTN 2004-503369.
FR thanks Merton College, Oxford, and the Greendale Scholarship for 
financial support.

%
\def\NPB#1#2#3{Nucl. Phys. {\bf B#1} #2 (#3)}
\def\PLB#1#2#3{Phys. Lett. {\bf B#1} #2 (#3)}
\def\PLBold#1#2#3{Phys. Lett. {\bf#1B} #2 (#3)}
\def\PRD#1#2#3{Phys. Rev. {\bf D#1} #2 (#3)}
\def\PRL#1#2#3{Phys. Rev. Lett. {\bf#1} #2 (#3)}
\def\PRT#1#2#3{Phys. Rep. {\bf#1} #2 (#3)}
\def\ARAA#1#2#3{Ann. Rev. Astron. Astrophys. {\bf#1} #2 (#3)}
\def\ARNP#1#2#3{Ann. Rev. Nucl. Part. Sci. {\bf#1} #2 (#3)}
\def\MPL#1#2#3{Mod. Phys. Lett. {\bf #1} #2 (#3)}
\def\ZPC#1#2#3{Zeit. f\"ur Physik {\bf C#1} #2 (#3)}
\def\APJ#1#2#3{Ap. J. {\bf #1} #2 (#3)}
\def\AP#1#2#3{{Ann. Phys. } {\bf #1} #2 (#3)}
\def\RMP#1#2#3{{Rev. Mod. Phys. } {\bf #1} #2 (#3)}
\def\CMP#1#2#3{{Comm. Math. Phys. } {\bf #1} #2 (#3)}
\relax
%
\newcommand{\journal}[4]{{ #1} {\bf #2}, #3 (#4)}
\newcommand{\hepth}[1]{{hep-th/#1}}
\newcommand{\hepph}[1]{{hep-ph/#1}}
\newcommand{\grqc}[1]{{gr-qc/#1}}
\newcommand{\astro}[1]{{astro-ph/#1}}
%

\end{document}